\documentclass[iop]{emulateapj}

\slugcomment{Accepted for Publication in ApJ}

\usepackage{color}
\usepackage{amsmath}
\usepackage{natbib}
\begin{document}

\title{Tracking the Evolution of QPO in RE J1034+396 Using the Hilbert-Huang Transform}

\author{Chin-Ping Hu, Yi Chou, Ting-Chang Yang, Yi-Hao Su}
\affil{Graduate Institute of Astronomy, National Central University, Jhongli 32001, Taiwan}
\email{m929011@astro.ncu.edu.tw, yichou@astro.ncu.edu.tw}

\begin{abstract}
RE J1034+396, a narrow-line Seyfert-1 active galactic nucleus (AGN), is the first example of AGNs that exhibited a nearly coherent quasi-periodic oscillation (QPO) for the data collected by XMM-Newton in 2007. The spectral behaviors and timing properties of the QPO have been studied since its discovery. We present an analysis of the QPO in RE J1034+396 based on the Hilbert-Huang transform (HHT). Comparing other time-frequency analysis methods, the Hilbert spectrum reveals the variation of the QPO period in great detail. Furthermore, the empirical mode decomposition provides band-pass filtered data that can be used in the O -- C and correlation analysis. We suggest that it is better to divide the evolution of the QPO in this observation into three epochs according to their different periodicities. In addition to the periodicities, the correlations between the QPO periods and corresponding mean count rates are also different in these three epochs. Further examining the phase lags in these epochs, we found no significant phase lags between the soft and hard X-ray bands, which is also confirmed in the QPO phase-resolved spectral analysis. Finally, we discuss the indications of current models including a spotted accretion disk, diskoseismology, and oscillation of shock, according to the observed time-frequency and spectral behaviors. 
\end{abstract}

\keywords{accretion disks --- galaxies: Seyfert --- galaxies: individual (RE J1034+396) ---X-rays: galaxies}

\section{Introduction}

Quasi-periodic oscillations (QPOs), which contain useful information about the inner accretion disk, are commonly observed in black hole X-ray binaries. Active galactic nuclei (AGNs), as upscaling of black hole X-ray binaries, are expected to have QPO phenomenon. However, AGNs were never significantly found to exhibit QPOs until the first detection in a narrow-line Seyfert 1 (NLS1) galaxy RE J1034+396 made by XMM-Newton observation \citep{Gierlinski2008}.  A QPO with period of 3733 s was found to be coherent with the data segment after $\sim25$ ks of this XMM-Newton observation at a $5.6 \sigma$ confidence level \citep{Gierlinski2008}.  From the observation on bulge stellar velocity dispersion, \citet{Bian2010} estimated that the mass of the supermassive black hole (SMBH) is $(1-4) \times 10^6 M_{\odot}$.  This estimation indicated that this QPO corresponds to a high-frequency QPO in the black hole X-ray binary \citep{Gierlinski2008, Middleton2009, Bian2010}.  

RE J1034+396, as well as other NLS1s, are also known for their strong soft X-ray excess and highly variability \citep{Puchnarewicz2001}.  The spectral behavior may also relate to the presence of QPO. Using the rms spectrum, \citet{Middleton2009} concluded that the energy spectrum of RE J1034+396 could be decomposed by a low-temperature Comptonization of disk emission and a hard power law tail. The QPO is dominated by the variability of the hard power law tail, while the soft component remains constant. Furthermore,  \citet{Middleton2011} folded the light curve of different energy bands with the QPO period and found a significant soft lag, which can be interpreted by the reprocessing. \citet{Maitra2010} divided the X-ray photons into high and low phases and found that the low phase spectrum exhibits an absorption edge at $\sim 0.86$ keV, which corresponds to a warm absorber lying at $\sim 9 r_g$ for a $4\times 10^6 M_{\odot}$ black hole.  

QPOs in X-ray binaries result in broad peaks in their power spectra. This broadening is probably caused by the dramatic variation in the QPO frequency or a modulation with a relatively stable period but fragmented in time. Fourier analysis is insufficient to study the time variability of frequencies because the frequencies defined in the Fourier spectra are assumed to be constant over the entire observation. Based on the timing analysis technique, adding a moving window is a straightforward solution to investigate the variation of the frequency. Several similar analysis methods were introduced, such as the dynamic power spectrum \citep{Clarkson2003a}, and spectrogram \citep{Oppenheim1989}.  An improved time-frequency analysis method, i.e., the Morlet wavelet analysis, was also invented and widely applied on astronomical time series, especially quasi-periodic modulations in various time scales. For example, \citet{Lachowicz2010} studied the 4 Hz low-frequency QPO in XTE J1550--564 using the wavelet analysis and Matching Pursuit algorithm. They concluded that the QPO is composed of multiple independent oscillations with the same frequencies, but the oscillation was present intermittently. For the case of RE J1034+396, \citet{Czerny2010} found a drift in the QPO central frequency based on the wavelet analysis. A possible $\sim24$ ks time scale of a QPO period variation was also marginally detected but could not be firmly concluded as a periodicity because the data time span was limited. In addition, they also found a positive correlation between the QPO frequencies, which were determined by the frequencies at the peak of wavelet spectrum, and the X-ray fluxes.

Another possible method to investigate the variation of QPO period is the Hilbert-Huang Transform (HHT) proposed by \citet{Huang1998}.  The HHT has been successfully applied on the superorbital modulation of SMC X-1 \citep{Hu2011}, 11-years sunspot variability \citep{Barnhart2011}, and the search for gravitational waves \citep{Camp2007}.  In the HHT, the instantaneous frequency, which is different from that in the Fourier analysis, is defined as the time derivative of phase function. Thus, the Hilbert spectrum can provide us detailed information in both the time and frequency domains. However, the time interval between the samplings must be much shorter than the variability time scale. The fast oscillations in X-ray binaries, such as the kilohertz QPOs and burst QPOs, cannot be analyzed by the HHT with current observatories. Fortunately, the X-ray intensity of RE J1034+396 is high enough to provide us with a sufficiently sampled X-ray light curve that can be analyzed using the HHT. In addition, HHT can provide us the phase function of the QPO. Thus, phase-resolved spectral analysis, as well as the phase lags between folded light curves of different energy bands are also applicable.

We present our analysis on the evolution of the QPO period for RE J1034+396 as well as the QPO phase-resolved spectral variation.  In Section \ref{obs}, we briefly introduce the observation made by XMM-Newton and the data selection criteria. The time-frequency analysis, including the HHT analysis, O -- C result, issues with phase lags, and relationship between the QPO period and X-ray flux, are presented in Section \ref{analysis}.  The spectral model as well as the phase-resolved spectral analysis are described in Section \ref{spectral}.  We further discuss different possible scenarios, including the spotted accretion disk model, diskoseismology, and shock oscillation model in Section \ref{discussion}.  Finally, we summarize our work in Section \ref{summary}. 

\section{Observation}\label{obs}
The QPO in RE J1034+396 was detected by a $\sim93$ ks XMM-Newton observation on May 31, 2007 (OBSID: 0506440101). Both the MOS and PN detector were operated in full-frame mode. Following the selection criteria of \citet{Gierlinski2008}, we extracted the photon events within 45'' radius around the source, and then excluded the final $\sim7$ ks data because they were highly affected by the background flare. All the photon arrival times were corrected to the barycenter of the solar system. We further combined the photons collected by MOS1, MOS2, and PN detectors with the energy range between 0.3 and 10 keV to perform a time-frequency analysis.

\section{Time-Frequency Analysis}\label{analysis}
\begin{figure*}
\epsscale{0.8}
\plotone{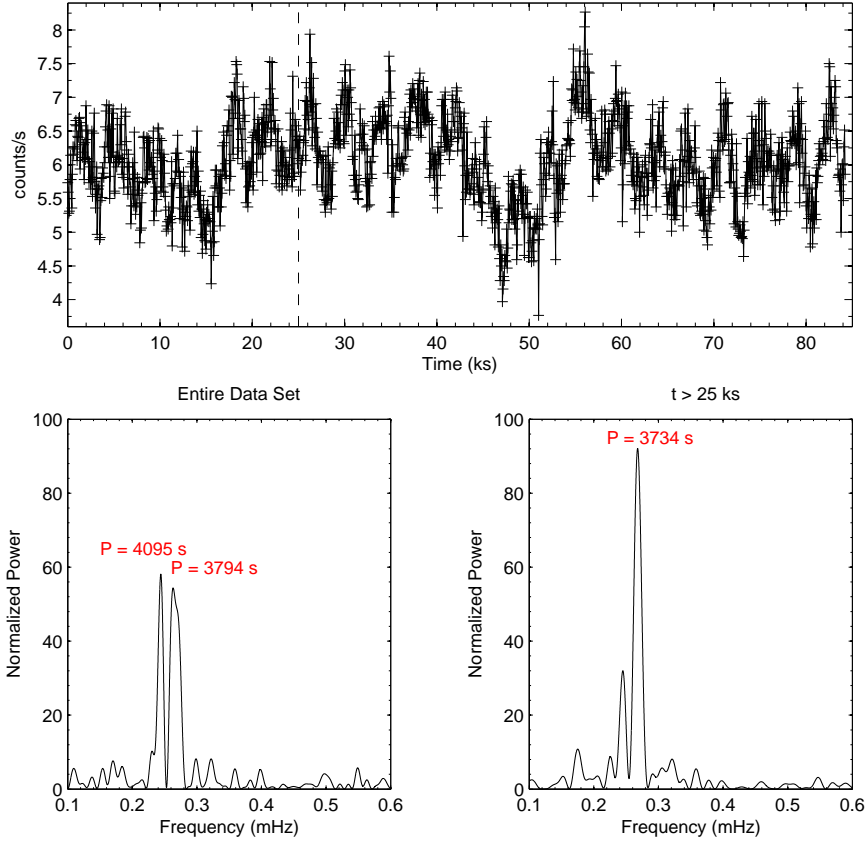} \caption{Upper: Merged X-ray light curve of RE J1034+396 collected by PN+MOS detectors on XMM-Newton. The dashed line denotes $t=25$ ks, which divided the light curve into two segments \citep[see][]{Gierlinski2008}.  Lower: The Lomb-Scargle periodogram of the entire data set (left) and of the segment 2 light curve (right). \label{LC_LS}}
\end{figure*}

In order to have sufficient samplings within one cycle, the X-ray photons were evenly binned into a 64-s resolution light curve shown as the upper panel of  \ref{LC_LS}.  As described in \citet{Gierlinski2008}, the QPO modulation is easily seen. A flux-drop event in which the count rate drops from $\sim6.5$ counts/s to $\sim5$ counts/s occurs between $t\approx40-55$ ks.  This event was noticed by \citet{Middleton2009} and believed to be an occultation. In addition, a similar event is also observed in $\sim10-20$ ks although it is less pronounced.

To check the consistency with the Fourier analysis performed by \citet{Gierlinski2008}, we applied the Lomb-Scargle periodogram \citep{Scargle1982, Press1989} on the entire light curve and segment 2 defined by \citet{Gierlinski2008}, i.e., after 25 ks. The power spectra are shown in the lower panel of Figure \ref{LC_LS}.  The spectrum of the segment 2 light curve exhibits a strong peak at $P=3733.6\pm3.9$ s, where the error comes from a $10^4$ times Monte Carlo simulation. This period is consistent with the one (3733 s) reported by \citet{Gierlinski2008}.  On the other hand, the power spectrum of the entire data set shows two significant peaks located at $P = 3794 \pm 19$ s and $P=4095 \pm 4$ s.  This suggests that the QPO period is probably not stable during the observation. The presence of these two peaks motivated us to further study its time-frequency properties using the HHT.  

\begin{figure*}
\epsscale{0.8}
\plotone{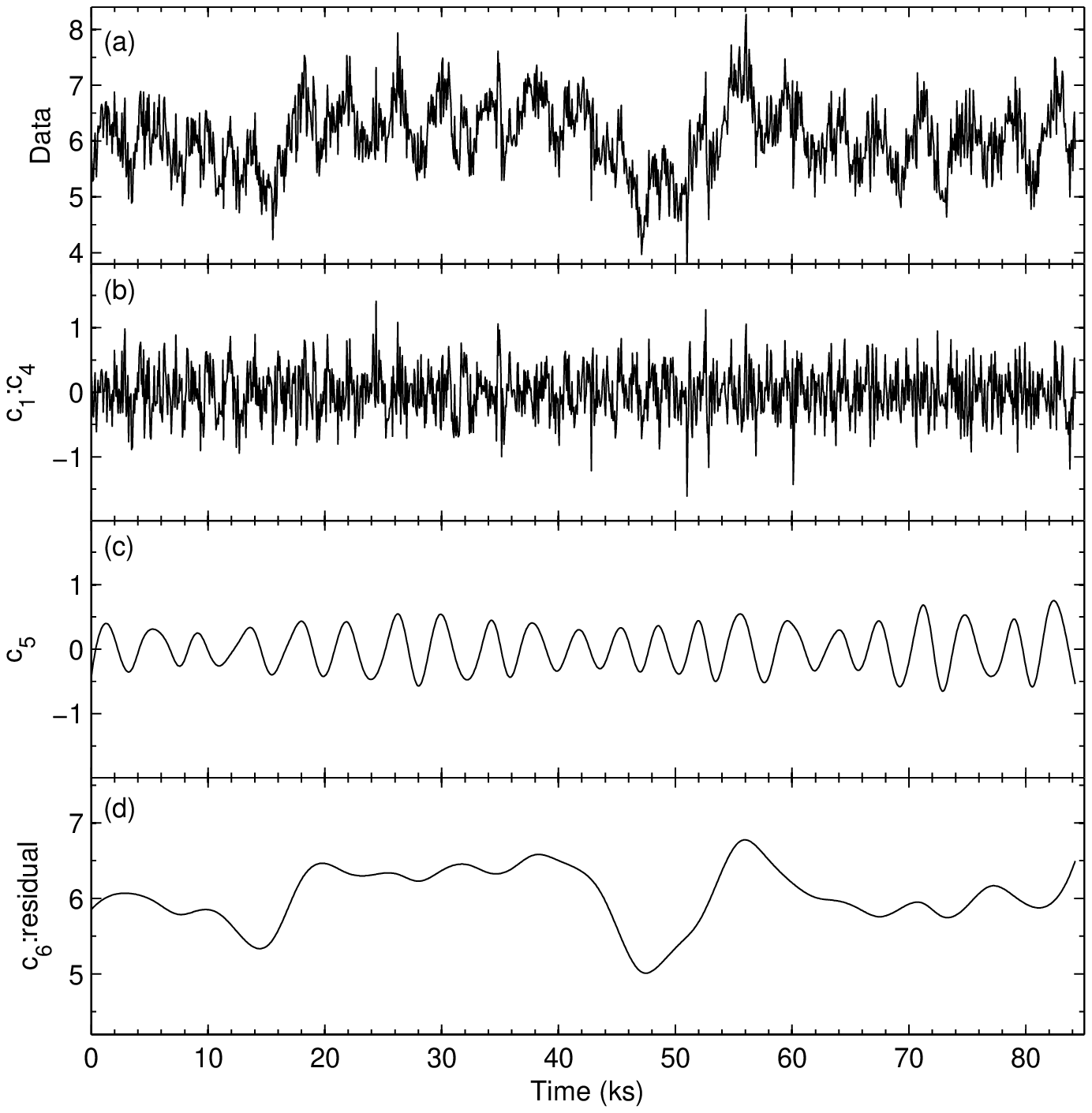} \caption{ (a) The original light curve. (b) EEMD high-pass filtered light curve, which is the summation of IMFs $c_1$ to $c_4$. (c) The fifth component, $c_5$, which corresponds to the QPO period. (d) The long-term modulation light curve, which is the summation of IMFs $c_6$ to $c_{10}$ and the residual, $r_{10}$.  Units of all the four light curves are counts/s. The completeness of EEMD guarantees that the summation of all the IMFs and residuals is the same as the original light curve.  \label{EEMD}}
\end{figure*}

\subsection{HHT Analysis}\label{HHT_analysis}
To obtain the variation of QPO frequency, we applied a modified version of the empirical mode decomposition (EMD), i.e., the ensemble EMD (EEMD) proposed by \citet{Wu2009}, on the combined PN+MOS light curve. The EMD is an iterative ``sifting'' process for extracting oscillation modes by subtracting the local means from the original data \citep{Huang1998}.  The decomposed components are intrinsic mode functions (IMFs) that satisfy the following two conditions: (1) the number of extrema and the number of zero crossings must be identical or differ by one; and (2) the local mean of the data is zero \citep{Huang1998, Hu2011}.  The instantaneous frequency obtained by the Hilbert transform is meaningful only if it is calculated from an IMF.  EEMD overcomes the mode-mixing problem, i.e., a modulation with the same time scale distributed across different IMFs, by taking the emsemble mean of the IMFs from combinations of the original data and added white noises \citep{Wu2009}.  Because the ensemble means of IMFs may not satisfied the condition of IMF, a post-processing EMD \citep{Wu2009} was applied on the decomposed components to guarantee that the final result satisfied the IMF criteria.

The light curve, $x(t)$, was decomposed into 10 IMFs, denoted as $c_1$ to $c_{10}$, and a residual, $r_{10}$.  The completeness of the IMFs are guaranteed so that the light curve can be represented as: 

\begin{equation}
x(t)=\sum_{j=1}^{10}c_j+r_{10}
\end{equation}
The QPO signal lies within $c_5$ can be examined by its variability time-scale as well as its timing behavior such as its Lomb-Scargle periodogram and dynamic power spectrum.  The orthogonality between neighboring components $c_f$ and $c_g$ can be expressed using the orthogonality index $OI_{fg}$ \citep{Huang1998}: 

\begin{equation}
OI_{fg}=\sum_{t}\frac{c_fc_g}{c_f^2+c_g^2}
\end{equation}
The orthogonality index between $c_4$ and $c_5$ in this decomposition is $OI_{45}=0.033$, whereas the orthogonality index between $c_5$ and $c_6$ is $OI_{56}=0.007$.  This means that the QPO signal is mostly concentrated within the IMF $c_5$, and the power barely leaks out to neighboring IMFs. Figure \ref{EEMD} shows the original light curve, EEMD high-pass filtered light curve, QPO component, and long-term modulation. The EEMD high-pass filtered light curve is a summation from $c_1$ to $c_4$, which represents the high-frequency noise in the data. The long-term modulation light curve, which represents the variability with time scale longer than the QPO, is a summation from $c_6$ to the residual that contains no extrema.  

We then applied the normalized Hilbert transform \citep{Huang2009} to the QPO component to obtain the instantaneous frequency and amplitude. The instantaneous amplitude $a_j(t)$ can be defined by the upper envelope of the absolute value of an IMF $c_j(t)$.  The Hilbert transform of a normalized IMF $X_j(t)=c_j(t)/a_j(t)$   can be represented as
\begin{equation}
Y(t)=\frac{1}{\pi}P\int_{-\infty}^{\infty}\frac{X(t')}{t-t'}dt'
\end{equation}
where P indicates the Cauchy principal value.  Therefore, an analytical signal $Z_j(t)$ as well as the angular phase function $\theta(t)$ can be defined as
\begin{equation}
Z_j(t)=X_j(t)+iY_j(t)=e^{i\theta_j(t)}
\end{equation}
Then, the instantaneous angular frequency $\omega_j(t)$ is defined as the time derivative of $\theta_j(t)$.  As a result, the data can be expressed as
\begin{equation}
x(t)=\sum_{j=1}^{10}a_j(t)\exp\left( i\int\omega_j(t)dt\right) +r_{10}
\end{equation}

\begin{figure}[h]
\epsscale{1.2}
\plotone{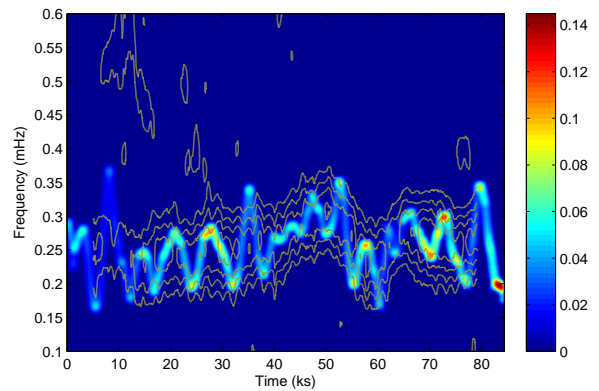} \caption{ The color map represents the Hilbert spectrum, where the color denotes the amplitude of the QPO. The contour is the dynamic Lomb-Scargle periodogram of the de-trended light curve. \label{HHT_DPS}}
\end{figure}

The variation of frequency and amplitude was plotted on a three-dimensional map, as shown in Figure \ref{HHT_DPS} to demonstrate the drift of the QPO frequency and the variation of the QPO amplitude. We also produced the dynamic Lomb-Scargle periodogram for comparison. The window size of the dynamic power spectrum was set to 10 ks and the moving step was 100 s. Instead of the original light curve, we applied the dynamic Lomb-Scargle algorithm on the detrended light curve, which was obtained by subtracting the long-term modulation light curve (Figure \ref{EEMD}(d)) from the original light curve. Because the window size is approximately the same as the time scale of flux-drop event $\sim40-55$ ks mentioned in \citet{Middleton2009}, the power spectra are dominated by the long-term trend, and the QPO signals are strongly depressed when the window moves around here. The dynamic power spectrum is shown as the contour stacked on top of the Hilbert spectrum in Figure \ref{HHT_DPS}.  

\begin{figure}[h]
\epsscale{1.2}
\plotone{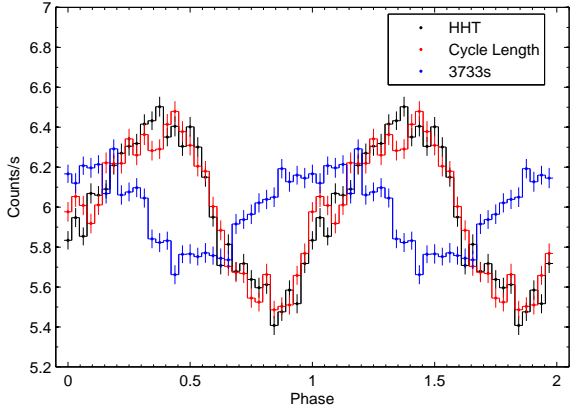} \caption{ The folded light curve folded by 3733 s (blue), HHT phase (black), and cycle length (red).   \label{fold_lc}}
\end{figure}

From Figure \ref{HHT_DPS}, we found that the Hilbert spectrum agrees with the dynamic power spectrum. Both of them can describe the drift of the QPO frequency for time scales longer than $\sim 10$ ks.  The flux-drop event, together with a frequency increase in $\sim40-55$ ks, can be observed in this time-frequency map. Moreover, the Hilbert spectrum shows more details in both the time and frequency domains than that in the dynamic power spectrum. The HHT provides us not only the instantaneous frequency and amplitude, but a well-defined phase function of QPO. We obtained a folded light curve by folding all the data according to the HHT phase.  Figure \ref{fold_lc} shows the light curves folded by the HHT phase, as well as ones folded by the cycle length and by a fixed period of 3733 s. The cycle length is the time difference between the neighboring fiducial points, which is defined by the local minima of the EEMD low-pass filtered light curve obtained from a summation of QPO signal and the long-term modulation light curve. This is similar to the definition of local minima in \citet{Trowbridge2007}, but we used the EEMD as a local filter instead of a Gaussian filter, and we used the local minima without fitting them with any functions. Because the QPO period is not stationary, the light curve folded with a fixed period of 3733 s would be expected to have distortion in its modulation shape. From Figure \ref{fold_lc}, the folded light curves produced by the HHT phase and cycle length are similar, whereas the one folded by a fix period is significantly different. We believe that the light curve folded by the HHT phase as well as that folded by the cycle length is a better description of the QPO profile than that folded by the fixed period. 

\subsection{O -- C Result}\label{o-c}
\begin{figure*}
\epsscale{0.8}
\plotone{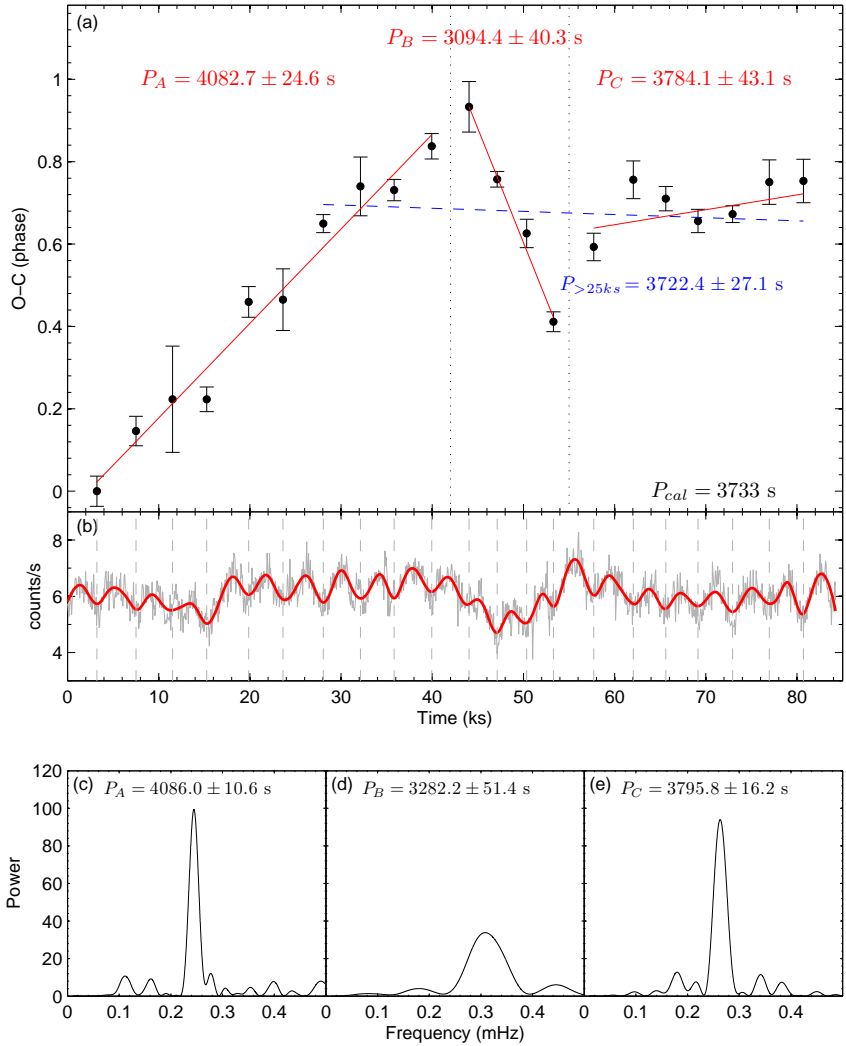} \caption{ (a) The O -- C result for QPO evolution.  Red lines are the linear fits for three different epochs; the blue dashed line is the linear fit of the segment 2 light curve. (b) The observed light curve (gray) and low-pass filtered light curve (red). The vertical gray dashed lines denote the minima determined by the filtered light curve.  (c)--(e) The corresponding Lomb-Scargle periodogram for epoch A-–C, respectively. \label{o_c}}
\end{figure*}

To study the evolution of the QPO period in detail, we applied the O -- C analysis on the QPO cycle length. The variation of QPO period in RE J1034+396 is so dramatic that it can be detected in both the dynamic power spectrum and Hilbert spectrum. However, it is still worth tracing the evolution of the QPO period using O -- C as long as the modulation is clearly seen without large observation gaps, which may cause the cycle count ambiguity. The fiducial point of each QPO cycle is defined as the one for the cycle length described in Section \ref{HHT_analysis}.  This low-passed filtered light curve provided us a model independent way to define the fiducial point, i.e., the minimum, in each cycle for the O -- C analysis.  We further applied a $10^4$ times Monte Carlo simulation to determine the error of the occurance time of fiducial points. Then, we applied a linear ephemeris with the location of the first minimum as the phase zero epoch, and the computed period was 3733 s. Figure \ref{o_c} (a) shows the phase evolution of QPO and Figure \ref{o_c} (b) shows the corresponding light curve, EEMD low-pass filtered light curve, and local minima. From the O -- C results, we found that it is better to divide the evolution of the QPO in three epochs rather than two segments proposed by \citet{Gierlinski2008}.  Epoch A is from the beginning of the observation to $\sim 42$ ks, and the corresponding period obtained by fitting the phase evolution with a straight line is $P_A=4082.7\pm24.6$ s. Epoch B starts at $\sim 42$ ks and ends at $\sim 55$ ks, which corresponds to the flux-drop event. The period of epoch B obtained by O -- C is $P_B=3094.4\pm40.3$ s, which much shorter than that of epoch A. This decrease in period is also observed in the time-frequency map (see Figure \ref{HHT_DPS}).  Because the flux-drop occurs with an increase in frequency, it strongly implies that this event is caused by the internal change in the QPO status rather than an external occultation as suggested by \citet{Middleton2009}.  Epoch C is $t\gtrsim55$ ks, and the period is derived as $P_C=3784.1\pm43.1$ s, which is also significantly shorter than that of epoch A.  The overall $\chi_{\nu}^2$ is 2.24, which implies a moderate fitting. Because the cycle length varies dramatically even between neighboring cycles, the large $\chi_{\nu}^2$ can be interpreted as resulting from the instability of the QPO period. We also fitted the segment 2 data with a straight line and found a period of $P(t>25ks)=3722.4\pm27.1$ s, which is consistent with the 3733 s determined by \citet{Gierlinski2008}.  However, the large $\chi_{\nu}^2$ value of 15.9 indicates an unacceptable fitting. 

The periodicities obtained by O -- C can be further examined by the Lomb-Scargle periodogram. The power spectra of the three epochs are shown in Figures \ref{o_c} (c)--(e).  The power spectrum of epoch A shows a strong peak at $P_A=4086.0\pm10.6$ s, whereas the one for epoch C also shows a strong peak at $P_C=3795.8\pm16.2$ s.  The periods determined for epoch A and C are consistent with the result obtained by O -- C.  These periods are also consistent with the peaks in the power spectra of the entire data set (Figure \ref{LC_LS}).  Furthermore, the periods of epoch A and C significantly deviate from each other at the $\sim5\sigma$ and $\sim10\sigma$ levels for the results obtained from the O -- C analysis and Lomb-Scargle periodograms, respectively. The periodogram of epoch B exhibits a weak broad peak at $P_B=3282.2\pm51.4$ s.  This period marginally agrees with that determined by O -- C at only the $2\sigma$ level.  This is acceptable because epoch B only consists of four cycles, and the QPO period changes between neighboring cycles.

\subsection{Phase Lag}\label{lag}
Time lags either between different X-ray bands or between X-ray, UV, and optical wavelengths, are intriguing issues of AGNs. Furthermore, time lags in the different time scales indicate different physical mechanisms. \citet{Middleton2011} found a soft lag in the folded QPO light curve. However, the QPO period was unstable, which implies that the phase lags are probably an artificial effect due to folding the non-stationary light curve with a fixed period, as we have shown in Section \ref{HHT_analysis}.  We investigated this issue with the help of the HHT and O -- C results. 

\begin{figure*}
\plotone{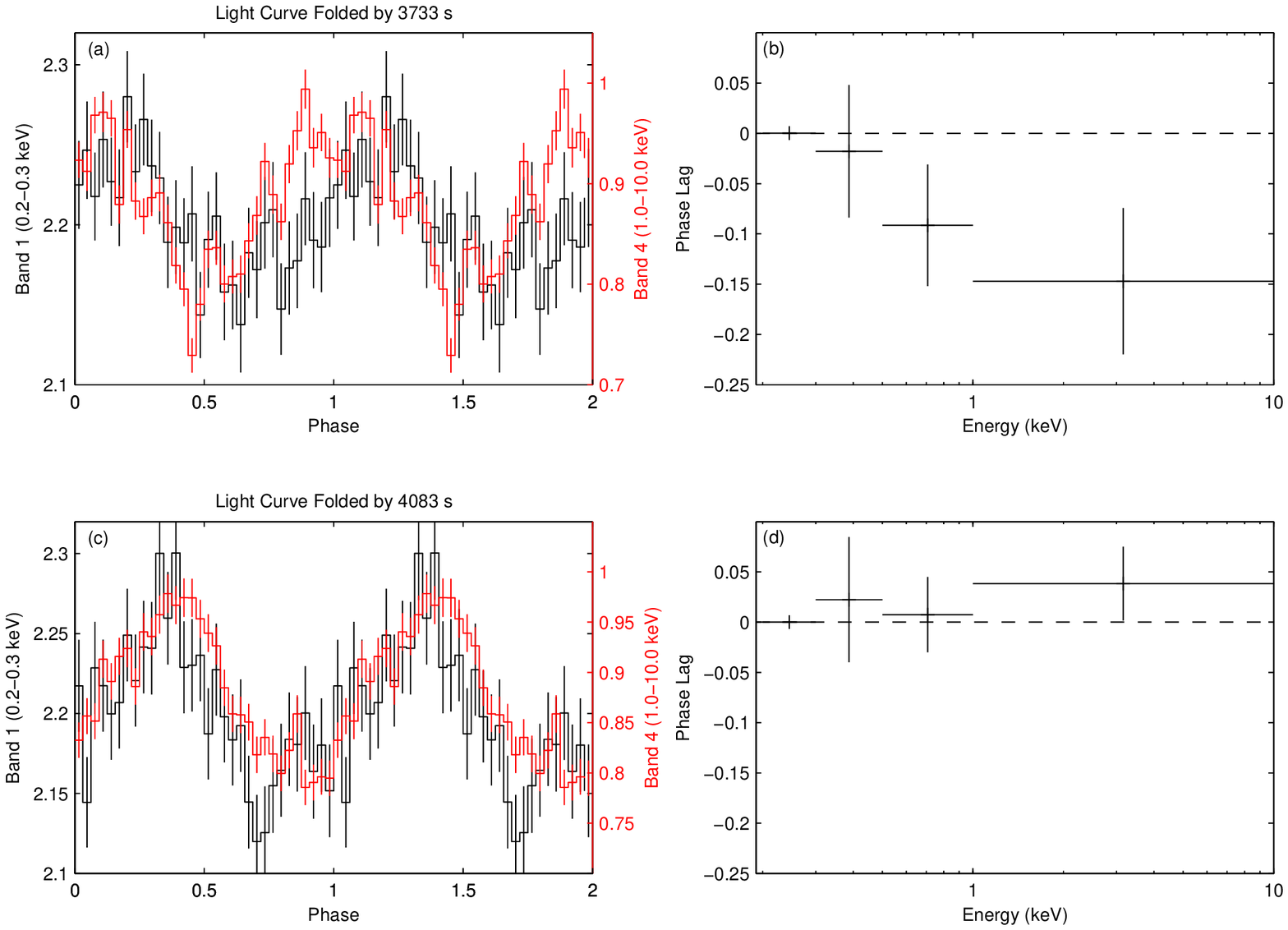} \caption{ (a) Folded light curves of two energy bands (0.2 -- 0.3 keV and 1.0 -- 10.0 keV).  The light curves were made by folding the entire data set by 3733 s. (b) The phase lags of the 0.3 -- 0.5, 0.5 -- 1.0, and 1.0 -- 10.0 keV bands with respect to the 0.2 -- 0.3 keV bands. (c) and (d) is the same as (a) and (b), respectively; however, the light curve is folded by 4083 s.  \label{phase_lag_3733_4083}}
\end{figure*}

\begin{figure*}
\plotone{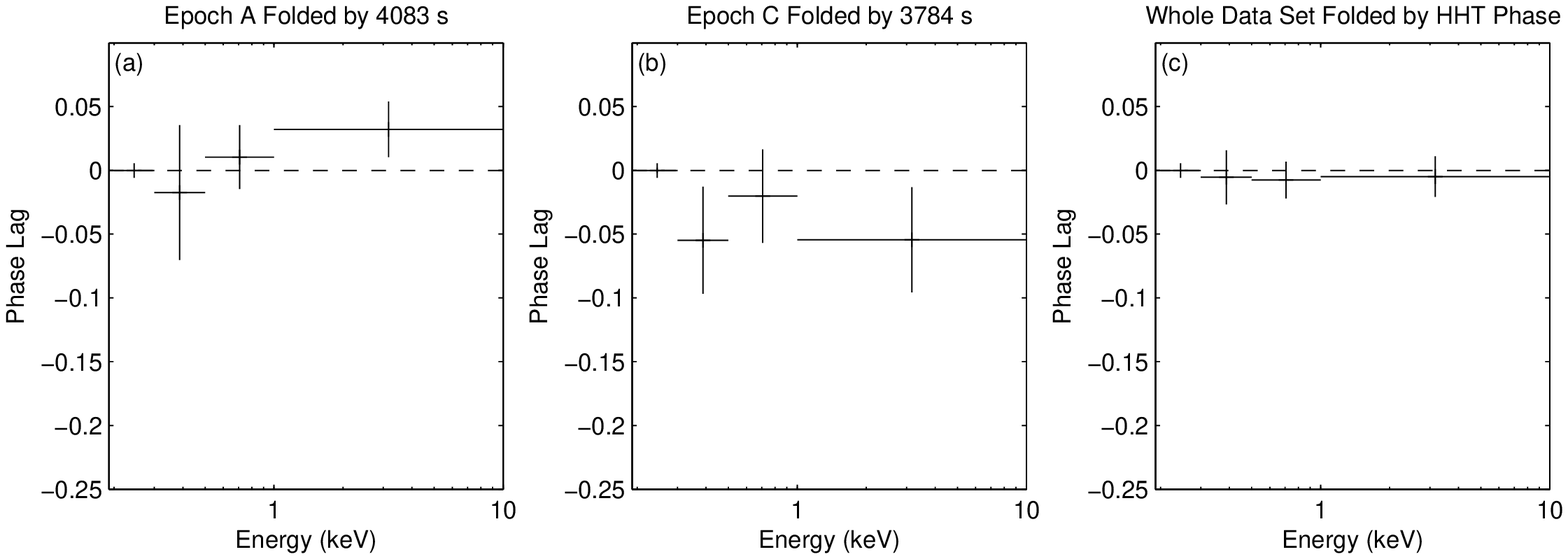} \caption{ (a) The phase-energy plot of the folded light curves of epoch A, where the folded period is 4083 s. (b) The same phase-energy plot of epoch C with a folded period of 3784 s. (c) The phase-energy plot of the folded light curves of the entire data set. The light curve were folded using the phase defined by the HHT.  \label{phase_lag_3784C_4083A_HHT}}
\end{figure*}

We first re-examined the results obtained in \citet{Middleton2011}.  The light curve was divided into four energy bands: band 1 (0.2 -- 0.3 keV), band 2 (0.3 -- 0.5 keV), band 3 (0.5 -- 1.0 keV), and band 4 (1.0 -- 10.0 keV). We then folded the light curves of these four energy bands using a period of 3733 s, and calculated the phase lags of bands 2 -- 4 with respect to band 1 using the cross-correlation method. The folded light curves with the largest phase lags, i.e., bands 1 and 4, are shown in Figure \ref{phase_lag_3733_4083}(a), where the soft lag can be easily seen. Figure \ref{phase_lag_3733_4083}(b) shows the relationship between the phase lags and energies, which is similar to that obtained by \citet{Middleton2011}.  However, from both the Lomb-Scargle periodogram and time-frequency analysis in the previous sections, we noticed that a single period of 3733 s is insufficient to describe the QPO behavior. For example, we also folded the light curve using a 4083 s period, which is another significant peak in the Lomb-Scargle periodogram in the entire data set. The folded light curves of bands 1 and 4 are shown as Figure \ref{phase_lag_3733_4083}(c), and the corresponding phase lags are shown as Figure \ref{phase_lag_3733_4083}(d).  No significant soft lag can be observed. 

Because the QPO period experienced three evolutionary epochs, we may obtain the phase lags between different energy bands of individual epochs. We first folded the light curves of four bands of epoch A using $P_A=4083$ s, and then observed the phase lags between the folded light curves. No significant phase lags were detected between bands 1 and 2 or bands 1 and 3 (Figure \ref{phase_lag_3784C_4083A_HHT}(a)).  The cross correlation analysis shows that the folded light curve in band 4 slightly lags band 1, but the significance is only at $\sim1\sigma$ level.  We processed similar analysis for epoch C by folding the light curve with period of $P_C=3784$ s.  Figure \ref{phase_lag_3784C_4083A_HHT}(b)  shows that the result is similar to that of epoch A, but band 4 appears to slightly lead band 1, although the significance is also at $\sim1\sigma$ level.  Finally, we folded the entire data set using the phase defined by the HHT to check the phase lags between different bands. However, no significant phase lags were detected (Figure \ref{phase_lag_3784C_4083A_HHT}(c)).

\subsection{Correlation between QPO Period and Flux}
\begin{figure}[h]
\epsscale{1.2}
\plotone{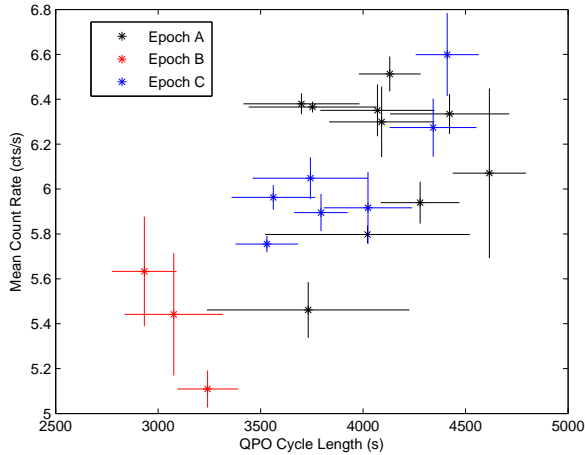} \caption{Relationship between the mean flux and QPO cycle length. The cycles of epoch A, B, and C are drawn in black, blue, and red, respectively.  \label{corr_mp_rate}}
\end{figure}

We then examined the relationship between the flux and QPO period for the entire data and 3 epochs, as defined in Section \ref{o-c}.  To obtain an independent measurement of the period and to avoid the influence of intra-wave frequency modulation, i.e., the frequency variation within one cycle, we used the cycle length obtained in Section \ref{o-c} as the period of each QPO cycle. On the other hand, we determined the flux of each QPO cycle by averaging the corresponding count rate of a long-term modulation light curve, which is shown in Figure \ref{EEMD}(d), to decrease the influence of the noise and QPO modulation. Twenty QPO cycles were used in the correlation analysis. The relationship between the mean flux and QPO period is shown in Figure \ref{corr_mp_rate}.  The linear correlation coefficient of the entire data set is 0.67 with a null hypothesis probability of $3.4\times10^{-4}$.   This indicates a strong correlation between the mean flux and QPO period. For the individual epochs, epoch B contains only 3 cycles, so we ignore it and focus on the other two epochs. The linear correlation coefficient of epoch A is 0.11 with a null hypothesis probability of 0.68, which indicates no significant correlation. In contrast, the linear correlation coefficient of epoch C is 0.85 with a null hypothesis probability of $7.4\times10^{-3}$, which is much stronger than that of epoch A. 

We further examined the slope between the QPO period and count rate to compare our results with that from \citet{Czerny2010}.  We made a logarithmic plot of the relationship between the mean count rate and QPO cycle length, and then fitted the data points of the entire data set with a straight line. The slope is $0.30\pm0.08$, which is marginally consistent with the results of \citet{Czerny2010} ($0.42\pm0.05$).  We calculated the slopes for epoch A and C to be $0.09\pm0.26$ and $0.43\pm0.12$, respectively.  The interpretation of the slopes as well as the dramatic change in the QPO period is discussed with the current QPO models in Section \ref{discussion}.  

\section{Spectral Analysis}\label{spectral}
From the fractional rms amplitude, \citet{Middleton2009} concluded that the spectral behavior of RE J1034+396 can be described as a constant low-temperature Comptonized component plus a highly variable power law component. \citet{Maitra2010} divided the QPO into high- and low-intensity phases, and then fitted the spectra using a blackbody component plus a broken power law component. They found an absorption feature by assuming that the spectral shape did not vary during the QPO cycle. To investigate the variation of spectral behavior during the QPO cycle, we performed a phase-resolved spectral analysis. 

Phase-resolved spectral analysis was achieved using the PN data because the PN collected more photons than the MOS detectors. The following selection criteria were used to achieve high quality spectra: pattern $\leq$ 4 and FLAG $=$ 0. The energy range was set between 0.2 to 10 keV. We first attempted to find the best spectral model to describe both the high-intensity and low-intensity phases. Then, the X-ray photons were divided into 16 phase bins to perform a phase-resolved spectral analysis with the best model. 

\subsection{Fits to High- and Low-Intensity Phases}\label{high_low}
To determine the best spectral model, we divided the QPO phase into a high-intensity phase (0.2 -- 0.55) and low-intensity phase (0.65--1.0) according to the folded light curve using the HHT phase. All the X-ray photons in the transition were discarded.  Because the image of RE J1034+396 was highly concentrated and had a heavy pile-up in the XMM-Newton PN and MOS chips \citep{Middleton2009, Maitra2010}, excising the X-ray photons within the central region is needed.  For the PN data of the entire data set, the $0.5-2$ keV observed-to-model fraction of single events after excising the inner $32"$ is $0.983\pm0.017$, which is acceptable at 1$\sigma$ level \citep{Maitra2010}.  We also used the SAS task {\tt epatplot} to examine this fraction of both the high- and low- intensity phases.  The observed-to-model fraction of single events is $0.986\pm0.014$ after excising the inner $15"$ region for the low-intensity state, whereas the fraction is $0.981\pm0.021$ after excising the inner $26"$ region for the high-intensity state. To reduce the statistical error, we grouped the X-ray photons using {\tt grppha} with a minimum of 100 photons in each energy bin. The {\tt XSPEC v12.8.0} of the {\tt HEASOFT} package was used to perform the spectral fitting. The galactic hydrogen column density was fixed at $1.31\times10^{20}$ cm$^{-2}$; however, an additional absorption to represent the absorption around the AGN was also set by adding a {\tt zphabs} with a fixed redshift of 0.042. We then used the spectral model proposed by \citet{Middleton2009}. A low-temperature Comptonized disk plus a hard power law tail (hereafter known as {\tt compTT+pl}) was used to fit both the high- and low-intensity phase data. Furthermore, we also added an additional blackbody component to the {\tt compTT+pl} as the seed photons, hereafter known as {\tt bb+compTT+pl}.  Table \ref{fitting_diff_model} lists the best-fit parameters of these two models for both the high- and low-intensity phases. 

\begin{table*}
\centering
\begin{tabular}{c|cc|cc}
\hline
Phase & \multicolumn{2}{c|}{High-intensity} & \multicolumn{2}{c}{Low-intensity} \tabularnewline
\hline
\hline 
Model & {\tt compTT+pl} & {\tt bb+compTT+pl} & {\tt compTT+pl} & {\tt bb+compTT+pl}\tabularnewline
\hline
$kT_{BB} (keV)$ & $-$ & $0.012\pm0.007$ & $-$ & $0.013\pm0.005$ \tabularnewline
$\Gamma$ & $1.71\pm1.1$ & $2.60\pm0.30$ & $1.91\pm0.52$ & $2.22\pm0.25$ \tabularnewline
$T_0 (keV)$ & $0.02\pm0.08$ & $=kT_{BB}$ & $0.02\pm0.06$ & $=kT_{BB}$ \tabularnewline
$kT_{comp} (keV)$ & $0.42\pm0.59$ & $0.16\pm0.04$ & $0.41\pm0.28$ & $0.24\pm0.05$ \tabularnewline
$\tau$ & $7.1\pm6.3$ & $13.8\pm3.3$ & $7.05\pm3.45$ & $10.6\pm1.6$ \tabularnewline
\hline
$\chi_{\nu}^{2}$ (dof) & 1.107(83) & 1.124(82) & 1.127(116) & 0.982(115) \tabularnewline
\hline
\end{tabular}
\caption{Best-fit parameters for high- and low-intensity phases. The $kT_{BB}$ is the blackbody temperature of blackbody model. For the Comptonized model, $T_0$ is the input seed photon temperature, $kT_{comp}$ is the plasma temperature, and $\tau$ is the plasma optical depth. The power law index of the power law tail is $\Gamma$. \label{fitting_diff_model}}
\end{table*}

The fitted results reveal a blackbody emission of $kT\sim10-\sim15$ eV, which probably represents the tail of the thermal emission from the gas in the inner accretion disk. Treating this blackbody emission as the seed photons for the {\tt bb+compTT+pl} model, we obtained a good fit for both the high- and low-intensity phases with $\chi_{\nu}^2\approx1.1$ and $1.0$, respectively.  The {\tt compTT+pl} model, by comparison, cannot constrain a meaningful soft photon temperature in both the high- and low-intensity phases. Thus, we chose the {\tt bb+compTT+pl} model to fit the phase-resolved spectra.

\subsection{Phase-Resolved Analysis}
We further divided the X-ray photons into 16 phase bins according to the HHT phase. All the X-ray photons collected in each phase bin were then fitted by the best-fit model in Section \ref{high_low}.  To increase the number of energy bins but keep sufficient statistics for each of the bins, especially the hard power law tail with energy higher than 2 keV, we grouped the X-ray photons with a minimum of 25 photons in each energy bin. All the spectral parameters were first set to be free, and we found that they all have no significant variation during the QPO cycle except for the normalizations. Then, we froze the spectral parameters to the values obtained in Section \ref{high_low}, except for the normalizations, to estimate the flux variation of the spectral components during one QPO cycle. Because of the lack of sufficient hard X-ray photons with energies higher than 6 keV in individual phase bins, we calculated the flux between 0.3 and 6.0 keV. The fluxes of black body component cannot be significantly detected because the normalization term of the black body component cannot be well constrained. 

\begin{figure*}
\plottwo{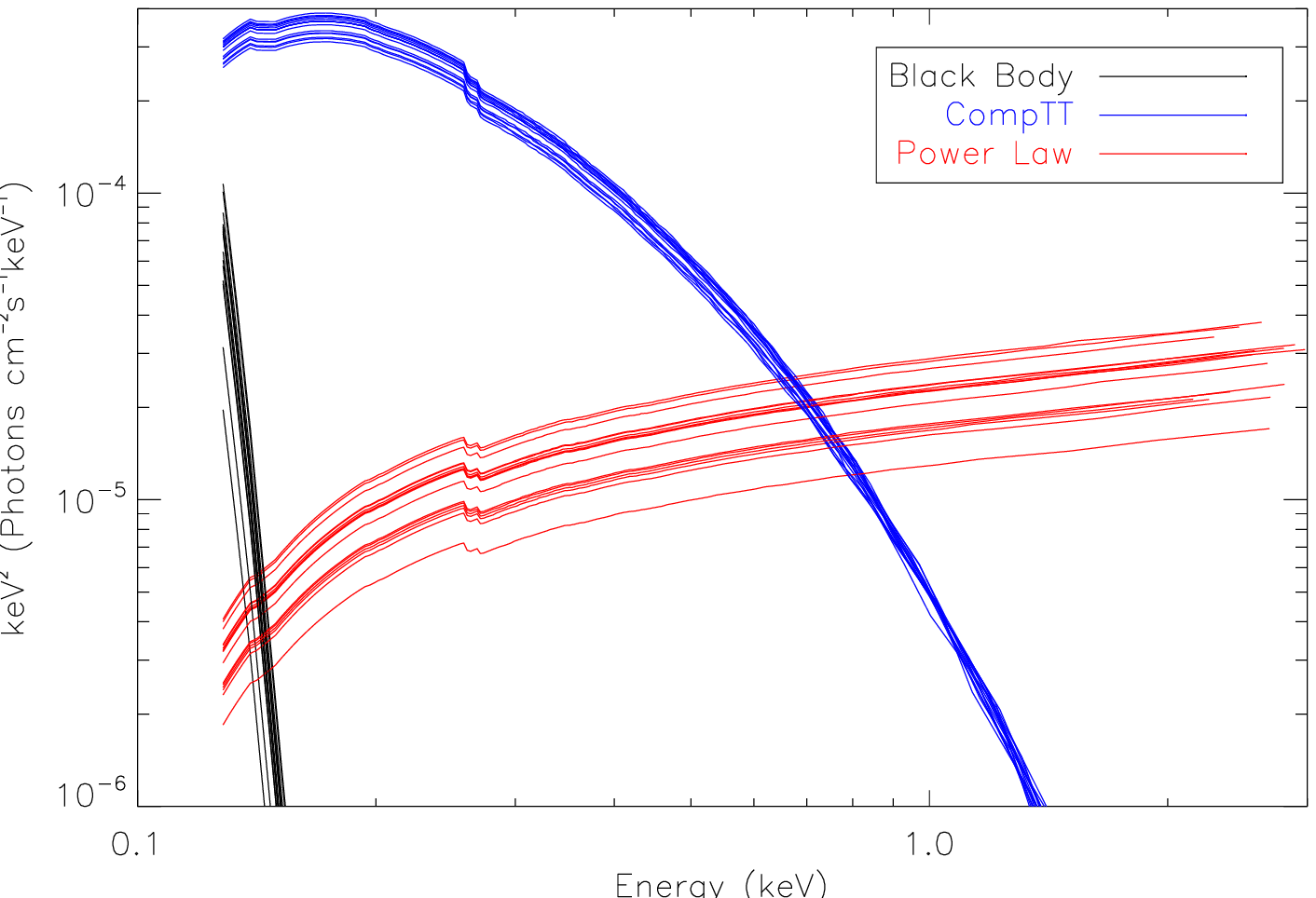}{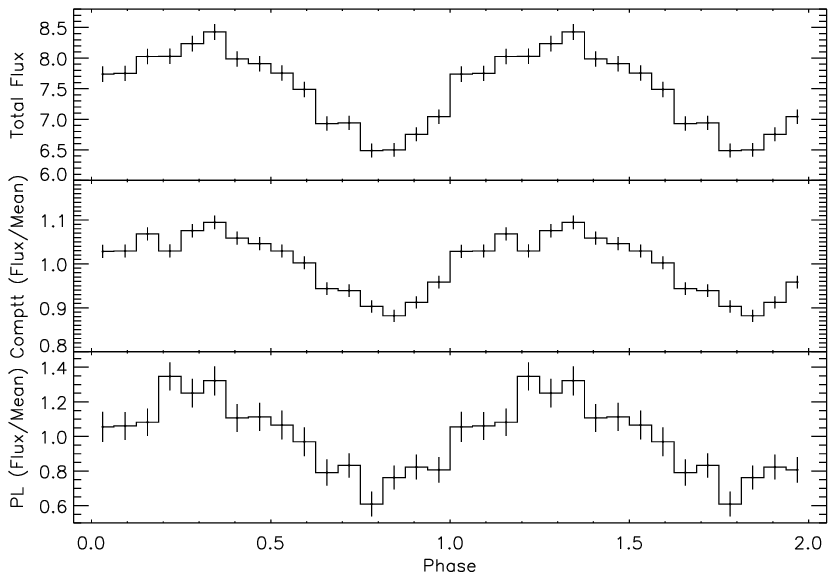} \caption{ Left: Best-fit models for 16 individual phase bins. Black, blue, and red lines represent the black body components, Comptonization of disk emissions, and highly variable power law tail, respectively. Right: The total flux and calculated model flux with respect to their means. The unit of flux is $10^{13}$ erg cm$^{-2}$ s$^{-1}$ \label{spectral_paremeter}}
\end{figure*}

The left panel of Figure \ref{spectral_paremeter} shows the variation of those three spectral components. The power law tail is more variable than the Comptonized component. We obtained the flux variation of these two components shown in the right panel of Figure \ref{spectral_paremeter}.  We found that the fractional rms amplitude of the power law component is 20.9\%, which is approximately three times larger than that of the Comptonized component (6.6 \%).  Furthermore, the phase lag between the Comptonized disk component and the power law tail is $-0.017\pm0.030$ where the error is estimated using a $10^4$ times Monte Carlo simulation, which indicates no significant phase lag. The lack of significant phase lag implies that if the Comptonized component is the reprocessed X-ray from the power law component, the light travel scale should be negligible  

\section{Discussion}\label{discussion}
The QPO detected in RE J1034+396 is believed to be a high-frequency QPO because of its short periodicity. This type of QPO in X-ray binary systems appears in the very high state, while the accretion disk nearly reaches the innermost region of the central black hole. A variety of mechanisms capable of interpreting the QPO phenomenon were proposed. We discussed three of them according to our discoveries, including the evolution of characteristic periods for each epoch, correlation between the QPO period and mean flux, and variation of spectral components.

\subsection{The Spotted Disk Model}
The most straightforward model to understand the QPO is that it is caused by the Keplerian motion of a temporary hot spot on the accretion disk \citep{Pechacek2006, Bachetti2010}. In this scenario, the QPO frequency is related to the radius of Keplerian orbit. From the HHT and O -- C analysis, we found that the QPO period changes dramatically even between neighboring cycles. This indicates that the hot spot is not strictly anchored on the accretion disk. Instead, the hot spot is wobbling around a characteristic radius. Furthermore, the evolution of the QPO period can be described by three epochs in this observation. This evolution can be interpreted as the change of characteristic radius caused by the instability of the accretion disk.

From the analysis of the relationship between modulation period and the flux, we observed a positive correlation between the QPO period and the flux.  This relationship, which is basically agree with that observed in \citet{Czerny2010}, does not agree with the prediction of this model.  However, some indications of this model are still worth to be addressed.  \citet{Czerny2010} compared the fractional rms amplitude of RE J1034+396 and that of the light curve calculated by a spotted accretion disk \citep{Pechacek2006}.  They concluded that the inclination angle should be 2 -- 3 degrees if the hot spot is close to the central black hole. However, not only the flare but also the disk, corona, and probably other materials near the black hole contribute to the X-ray emissions. For example, we obtained three components in the spectral fitting. Those X-ray sources would contribute non-modulated or lower-amplitude X-ray emissions such that the fractional amplitude as well as the inclination angle would be under estimated.  

\begin{figure}[h]
\epsscale{1.2}
\plotone{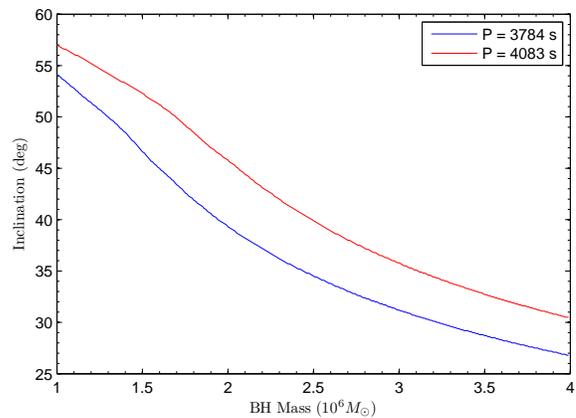} \caption{Estimated inclination angles for different black hole masses and QPO periods. The blue and red curves are the cases for QPO period of 3784 and 4083 s, respectively. \label{inclination_mass}}
\end{figure}

Besides the fractional amplitude, another possible indicator that can be used to estimate the inclination angle is the modulation shape. From the simulated light curve in \citet{Czerny2010} and \citet{Pechacek2006}, the QPO modulation shape became highly non-sinusoidal for higher inclination angles. A non-sinusoidal light curve can be expressed as harmonics in the Fourier analysis; however, we could not observe any significant harmonics in the power spectrum. There are two possible explanations for this. The first is that the inclination angle of RE J1034+396 is low, and the QPO profile is close to sinusoidal such that the harmonics are not significantly detected. Another possibility is that the QPO period varies so dramatically that the harmonics are more strongly depressed than the fundamental. Fortunately, the HHT provides us an indicator to probe the non-sinusoidal level of the data. Unlike harmonics in a Fourier analysis, the non-sinusoidal light curve would produce {\it intra-wave} frequency modulations, i.e., the frequency variation within one cycle \citep{Huang1998} in the instantaneous frequency. The frequency derived from the cycle length in the O -- C analysis can be treated as the {\it inter-wave} frequency modulation, i.e., the frequency variation between cycles. Thus, the intra-wave frequency modulation can be estimated by taking the difference between the instantaneous frequency and inter-wave frequency modulation. We then created simulated light curves according to equations 4 and 10 in \citet{Pechacek2006} for different black hole masses and inclination angles. The orbital radius of the hot spot is calculated using Kepler’s law for orbital periods of 3784 and 4083 s. The amplitude of the intra-wave frequency modulation of a simulated light curve can also be calculated using the Hilbert transform on the simulated light curve. For a mass range $1-4\times10^6$ M$_{\odot}$, we can estimate the inclination angle by matching the intra-wave frequency modulation amplitude of the simulated light curves with the observed one.  Figure \ref{inclination_mass} shows the relationship between the estimated inclination angle and black hole mass. We found that the inclination angle was between $\sim 27$ and $\sim 57$ degrees.  This estimation of inclination angle is probably an over-estimate because the noise in the light curve would also contribute to the intra-wave modulation.

\subsection{Diskoseismology}
Relativistic diskoseismology, i.e., the oscillations that are trapped in the inner accretion disk, may also be responsible for the presence of QPOs \citep{Perez1997, Silbergleit2001, Wagoner2001, Ortega2002}.  Three oscillation modes, i.e., g-, c-, and p-modes, are investigated. The fundamental frequency of these three modes mainly depend on the mass of the black hole, i.e., $f\propto1/M_{BH}$, where $f$ is the QPO frequency and $M_{BH}$ is the mass of black hole. From \citet{Wagoner2001}, the g-mode is dominated by the gravitational-centrifugal force, in which the oscillation frequency anti-correlates with the flux of accretion disk. The c-mode is caused by the non-radial corrugation on the inner accretion disk, while the oscillation frequency is highly dependent on the spin of black hole. The p-mode is dominated by the restoring-force due to the pressure gradient, in which the oscillation frequency shows a positive correlation with the flux of disk. Only the p-mode shows a positive correlation between the flux and QPO frequency, but it is opposite of our results. Thus, the QPO in RE J1034+396 is not likely caused by the p-mode oscillation. For a black hole with mass of 10 M$_{\odot}$, the g-mode oscillation is between $\sim 70$ to $\sim 110$ Hz for different angular momentums and luminosities \citep[see Figure 1 of][]{Wagoner2001}. If we enlarge the black hole mass to that in RE J1034+396, the g-mode oscillation period is between $\sim900$ and $\sim1400$ s for a $1\times10^6$ M$_{\odot}$ black hole, or between $\sim3600$ and $\sim5700$ s for a $4\times10^6$ M$_{\odot}$ black hole.  Thus, the preferred black hole mass is close to the upper limit of the estimation made by \citet{Bian2010}  if we assume that the QPO is caused by the g-mode oscillation. However, it is difficult for the slope between the g-mode oscillation period and luminosity to exceed $\sim0.1$ unless the angular momentum of the black hole is extremely high \citep[see equation 3 in][]{Wagoner2001}.Thus, the g-mode oscillation cannot provide the observed slope ($\sim0.3$), although this value is less reliable relative to the correlation coefficient because the observed X-ray flux may also contain the emissions from other components. On the other hand, the c-mode oscillation can produce a larger slope between the flux and QPO frequency in the range of our observed value, but the angular momentum of the black hole should be close to zero \citep[see equation 6 in][]{Wagoner2001}. However, the fundamental c-mode frequency of such a slow-rotating black hole is very low. For example, a 10 M$_{\odot}$ black hole with angular momentum of $a=0.05$ has a fundamental c-mode oscillation frequency of $f\lesssim1$ Hz.  This oscillation corresponds to a period longer than $10^5$ s for a $10^6$ M$_{\odot}$ black hole. Thus, if the c-mode oscillation is responsible for the observed QPO, it should be high-order harmonics.

We also studied the relationship between the QPO frequency and flux in detail by dividing the evolution of the QPO into three epochs. A change in correlation between epochs was observed: a significant anti-correlation between the frequency and flux was observed in epoch C, but no significant correlation was found in epoch A. This is possibly caused by a change in oscillation mode. A correlation in epoch A was not significantly detected; however, we found that the first three cycles appear to behave similar to the flux drop event in epoch B. If we ignore these three cycle, the remaining cycles in epoch A show a strong anti-correlation between the cycle length and flux with a linear correlation coefficient of $-0.61$.  Thus, the QPO in epoch A is probably caused by the p-mode oscillation.  After epoch B, i.e., the flux-drop event, the g-mode oscillation dominated the observed QPO phenomenon such that we observed a significant direct correlation between the cycle length and flux.

All the discussions above are based on the assumption that the luminosity is less than the Eddington luminosity. This is a necessary condition for the thin-disk assumption. However, the luminosity of RE J1034+396 is probably a super-Eddington \citep{Gierlinski2008, Middleton2009}.  Thus, we could not exclude all the possible oscillation modes in a super-Eddington accretion disk.

\subsection{Oscillation of Shock}
\citet{Das2011} proposed a novel model of QPO to investigate the observed slope between the QPO period and X-ray flux in RE J1034+396. In their model, QPO is caused by the oscillations of the shock formed inside the hot accretion flow. The variation in the QPO period is interpreted as the drifting of shock location. This model successfully interpreted the correlation between the QPO period and X-ray flux in \citet{Czerny2010} with slope of $0.92\pm0.03$.  However, we have demonstrated that the slope changes as QPO evolves in this observation. In fact, even the relationship likely changes. For example, the QPO period and flux exhibit no significant correlation in epoch A, or even anti-correlation when we omit the first three cycles that show another possible flux-drop event. Thus, the entire observation cannot be interpreted by this single scenario.

\section{Summary}\label{summary}
This research successfully applied the novel time-frequency analysis technique, i.e., HHT, on the QPO of RE J1034+396. The Hilbert spectrum enabled us to trace the variation of QPO period even between neighboring cycles. The EEMD filtered light curve allowed us to directly determine the fiducial points for the O -- C analysis. From the O -- C analysis, we found that the variation in the QPO period can be described as three evolutionary epochs. For a spotted accretion disk model and the shock oscillation model, the change of characteristic periods indicates a dramatic change in the flare or shock location, respectively. For the diskoseismology scenario, the change in QPO characteristic period as well as the change in flux-period correlation between epochs can be interpreted as the change in oscillation modes.

We further examined the phase lag phenomenon between soft and hard X-ray bands. No phase lags were determined when folding the light curve in different epochs with their corresponding periodicities, or folding the entire data set by the HHT phase. The phase lag determined by folding the entire data set with a fixed 3733 s period is probably an artificial effect that resulted from folding the variable periodicity with a fixed period. The lack of phase lags indicates that the emission regions for both spectral components are close to each other.

The time scale of QPO in AGN is much longer than that in the X-ray binary system. Such long time scales make it easier to study the variability in the QPO period in detail. We believe that future X-ray observations can discover more QPO samples in AGNs. With the help of the HHT and dynamic power spectrum, it is possible to study the evolution of QPO periodicities systematically.

\acknowledgments

We thank Prof. Christopher Reynolds and Prof. Chris Done for useful discussions in the QPO mechanisms.  We thank the anonymous referee for useful suggestions and comments. The HHT codes were provided by the Research Center for Adaptive Data Analysis in National Central University of Taiwan. This work is based on observations obtained with XMM-Newton, an ESA science mission with instruments and contributions directly funded by ESA Member States and the USA (NASA). This research was supported by the NSC 102-2112-M-008-020-MY3 grant from the Ministry of Science and Technology of Taiwan.

\bibliography{reference}

\begin{thebibliography}{}
\expandafter\ifx\csname natexlab\endcsname\relax\def\natexlab#1{#1}\fi

\bibitem[{{Bachetti} {et~al.}(2010){Bachetti}, {Romanova}, {Kulkarni},
  {Burderi}, \& {di Salvo}}]{Bachetti2010}
{Bachetti}, M., {Romanova}, M.~M., {Kulkarni}, A., {Burderi}, L., \& {di
  Salvo}, T. 2010, \mnras, 403, 1193

\bibitem[{{Barnhart} \& {Eichinger}(2011)}]{Barnhart2011}
{Barnhart}, B.~L., \& {Eichinger}, W.~E. 2011, \solphys, 269, 439

\bibitem[{{Bian} \& {Huang}(2010)}]{Bian2010}
{Bian}, W.-H., \& {Huang}, K. 2010, \mnras, 401, 507

\bibitem[{{Camp} {et~al.}(2007){Camp}, {Cannizzo}, \& {Numata}}]{Camp2007}
{Camp}, J.~B., {Cannizzo}, J.~K., \& {Numata}, K. 2007, \prd, 75, 061101

\bibitem[{{Clarkson} {et~al.}(2003){Clarkson}, {Charles}, {Coe}, {Laycock},
  {Tout}, \& {Wilson}}]{Clarkson2003a}
{Clarkson}, W.~I., {Charles}, P.~A., {Coe}, M.~J., {et~al.} 2003, \mnras, 339,
  447

\bibitem[{{Czerny} {et~al.}(2010){Czerny}, {Lachowicz}, {Dov{\v c}iak},
  {Karas}, {Pech{\'a}{\v c}ek}, \& {Das}}]{Czerny2010}
{Czerny}, B., {Lachowicz}, P., {Dov{\v c}iak}, M., {et~al.} 2010, \aap, 524,
  A26

\bibitem[{{Das} \& {Czerny}(2011)}]{Das2011}
{Das}, T.~K., \& {Czerny}, B. 2011, \mnras, 414, 627

\bibitem[{Gierl\'{i}nski {et~al.}(2008)Gierl\'{i}nski, Middleton, Ward, Done,
  {Gierli{\'n}ski}, {Middleton}, {Ward}, \& {Done}}]{Gierlinski2008}
Gierl\'{i}nski, M., Middleton, M., Ward, M., {et~al.} 2008, \nat, 455, 369

\bibitem[{{Hu} {et~al.}(2011){Hu}, {Chou}, {Wu}, {Yang}, \& {Su}}]{Hu2011}
{Hu}, C.-P., {Chou}, Y., {Wu}, M.-C., {Yang}, T.-C., \& {Su}, Y.-H. 2011, \apj,
  740, 67

\bibitem[{{H}uang {et~al.}(2009){H}uang, {Wu}, {Long}, {Arnold }, {Chen}, \&
  {Blank}}]{Huang2009}
{H}uang, N.~E., {Wu}, Z., {Long}, S.~R., {et~al.} 2009, Advances in Adaptive
  Data Analysis, 01, 177

\bibitem[{{Huang} {et~al.}(1998){Huang}, {Shen}, {Long}, {Wu}, {Shih}, {Zheng},
  {Yen}, {Tung}, \& {Liu}}]{Huang1998}
{Huang}, N.~E., {Shen}, Z., {Long}, S.~R., {et~al.} 1998, Royal Society of
  London Proceedings Series A, 454, 903

\bibitem[{{Lachowicz} \& {Done}(2010)}]{Lachowicz2010}
{Lachowicz}, P., \& {Done}, C. 2010, \aap, 515, A65

\bibitem[{{Maitra} \& {Miller}(2010)}]{Maitra2010}
{Maitra}, D., \& {Miller}, J.~M. 2010, \apj, 718, 551

\bibitem[{{Middleton} {et~al.}(2009){Middleton}, {Done}, {Ward},
  {Gierli{\'n}ski}, \& {Schurch}}]{Middleton2009}
{Middleton}, M., {Done}, C., {Ward}, M., {Gierli{\'n}ski}, M., \& {Schurch}, N.
  2009, \mnras, 394, 250

\bibitem[{{Middleton} {et~al.}(2011){Middleton}, {Sutton}, \&
  {Roberts}}]{Middleton2011}
{Middleton}, M.~J., {Sutton}, A.~D., \& {Roberts}, T.~P. 2011, \mnras, 417, 464

\bibitem[{{Oppenheim} \& {Schafer}(1989)}]{Oppenheim1989}
{Oppenheim}, A.~V., \& {Schafer}, R.~W. 1989, Discrete-Time Signal Processing
  (Prentice Hall)

\bibitem[{{Ortega-Rodr{\'{\i}}guez} {et~al.}(2002){Ortega-Rodr{\'{\i}}guez},
  {Silbergleit}, \& {Wagoner}}]{Ortega2002}
{Ortega-Rodr{\'{\i}}guez}, M., {Silbergleit}, A.~S., \& {Wagoner}, R.~V. 2002,
  \apj, 567, 1043

\bibitem[{{Pech{\'a}{\v c}ek} {et~al.}(2006){Pech{\'a}{\v c}ek}, {Dov{\v
  c}iak}, \& {Karas}}]{Pechacek2006}
{Pech{\'a}{\v c}ek}, T., {Dov{\v c}iak}, M., \& {Karas}, V. 2006, Astronomische
  Nachrichten, 327, 957

\bibitem[{{Perez} {et~al.}(1997){Perez}, {Silbergleit}, {Wagoner}, \&
  {Lehr}}]{Perez1997}
{Perez}, C.~A., {Silbergleit}, A.~S., {Wagoner}, R.~V., \& {Lehr}, D.~E. 1997,
  \apj, 476, 589

\bibitem[{{Press} \& {Rybicki}(1989)}]{Press1989}
{Press}, W.~H., \& {Rybicki}, G.~B. 1989, \apj, 338, 277

\bibitem[{{Puchnarewicz} {et~al.}(2001){Puchnarewicz}, {Mason},
  {Siemiginowska}, {Fruscione}, {Comastri}, {Fiore}, \&
  {Cagnoni}}]{Puchnarewicz2001}
{Puchnarewicz}, E.~M., {Mason}, K.~O., {Siemiginowska}, A., {et~al.} 2001,
  \apj, 550, 644

\bibitem[{{Scargle}(1982)}]{Scargle1982}
{Scargle}, J.~D. 1982, \apj, 263, 835

\bibitem[{{Silbergleit} {et~al.}(2001){Silbergleit}, {Wagoner}, \&
  {Ortega-Rodr{\'{\i}}guez}}]{Silbergleit2001}
{Silbergleit}, A.~S., {Wagoner}, R.~V., \& {Ortega-Rodr{\'{\i}}guez}, M. 2001,
  \apj, 548, 335

\bibitem[{{Trowbridge} {et~al.}(2007){Trowbridge}, {Nowak}, \&
  {Wilms}}]{Trowbridge2007}
{Trowbridge}, S., {Nowak}, M.~A., \& {Wilms}, J. 2007, \apj, 670, 624

\bibitem[{{Wagoner} {et~al.}(2001){Wagoner}, {Silbergleit}, \&
  {Ortega-Rodr{\'{\i}}guez}}]{Wagoner2001}
{Wagoner}, R.~V., {Silbergleit}, A.~S., \& {Ortega-Rodr{\'{\i}}guez}, M. 2001,
  \apjl, 559, L25

\bibitem[{Wu \& Huang(2009)}]{Wu2009}
Wu, Z., \& Huang, N.~E. 2009, Advances in Adaptive Data Analysis, 01, 1

\end{thebibliography}

\end{document}